\documentclass[%
 aip,
cp,  
 amsmath,amssymb,
 reprint,%
]{revtex4-2}

\usepackage{graphicx}
\usepackage{dcolumn}
\usepackage{bm}

\usepackage[utf8]{inputenc}
\usepackage[T1]{fontenc}
\usepackage{mathptmx}

\renewcommand{\exp}[1]{e^{#1}}

\newcommand{\w}{\omega}

\newcommand{\B}{\beta}

\renewcommand{\r}{\rho}

\newcommand{\Ap}{A_p}
\newcommand{\bp}{b_p}
\newcommand{\Bu}{B_u}

\newcommand{\pconj}{\overline{p}}

\renewcommand{\xi}{x_i}

\newcommand{\appropto}{\mathrel{\vcenter{
  \offinterlineskip\halign{\hfil$##$\cr
    \propto\cr\noalign{\kern2pt}\sim\cr\noalign{\kern-2pt}}}}}

\newcommand{\kx}{k}

\renewcommand{\Re}{\text{Re}}
\renewcommand{\Im}{\text{Im}}

\usepackage{xcolor}

\usepackage{longtable}
\usepackage{subfigure}

\begin{document}


\title{Cochlear Wave Propagation and Dynamics in the Human Base and Apex \\ [1ex] \large Model-Based Estimates from Noninvasive Measurements}

\author{Samiya A Alkhairy} 
 \email[Corresponding author: ]{samiya@mit.edu; samiya@alum.mit.edu}

\affiliation{
Massachusetts Institute of Technology, USA
}


\begin{abstract}

Cochlear wavenumber and impedance are mechanistic variables that encode information regarding how the cochlea works - specifically wave propagation and Organ of Corti dynamics. These mechanistic variables underlie interesting features of cochlear signal processing such as its place-based wavelet analyzers, dispersivity and high-gain. Consequently, it is of interest to estimate these mechanistic variables in various species (particularly humans) and across various locations along the length of the cochlea.
In this paper, we (1) develop methods to estimate the mechanistic variables (wavenumber and impedance) from noninvasive response characteristics (such as the quality factors of psychophysical tuning curves) using an existing analytic shortwave single-partition model of the mammalian cochlea. 
The model we leverage in developing the estimation methods is valid at low stimulus levels, was derived using a physical-phenomenological approach, and tested using a variety of datasets from multiple locations and species. The model’s small number of parameters and simple closed-form expressions enable us to develop methods for estimating mechanistic variables from noninvasive response characteristics. Developing these estimation methods involves (1a) deriving expressions for model constants, which parameterize the model expressions for wavenumber and impedance, in terms of characteristics of response variables - e.g. bandwidths and group delays of pressure across the Organ of Corti; followed by (1b) deriving expressions for the model constants in terms of noninvasive response characteristics. Using these derived expressions, we can estimate the wavenumber and impedance from noninvasive response characteristics for various species and locations along the length of the cochlea.
After developing the estimation methods, we (2) apply these methods to estimate human mechanistic variables, and (3) make comparisons between the base and apex. 
We estimate the mechanistic variables in the human base and apex following the methods developed in this paper and using reported values for quality factors from psychophysical tuning curves and a location-invariant ratio extrapolated from chinchilla. Our resultant estimates for human wavenumbers and impedances show that the minimum wavelength (which occurs at the peak of the traveling wave) is smaller in base than the apex. The Organ of Corti is stiffness dominated rather than mass dominated, and there is negative effective damping prior to the peak followed by positive effective damping. The effective stiffness, and positive and negative effective damping are greater in the base than the apex. 
Future work involves studying the closed-form expressions for wavenumber and impedance for qualitative mechanistic interpretations across mammalian species as well as studying derived mechanisms such as power flux into the traveling wave and features of the cochlear amplifier. The methods introduced here for estimating mechanistic variables from characteristics of invasive or noninvasive responses enable us to derive such estimates across various species and locations where the responses are describable by sharp filters. In addition to studying cochlear wave propagation and dynamics, the estimation methods developed here are also useful for auditory filter design.

\end{abstract}

\maketitle

\section{Motivation and objectives}

The cochlea has fascinating signal processing features which motivates our interest in understanding how it works. Stapes vibration results in traveling waves that propagate along the length of the cochlea and are subject to dispersion and amplification. Relative to the stapes stimulus, the response at each location along the length of the cochlea peaks at a particular frequency. Such features are of particular interest to auditory physicists as well as those interested in bio-mimetic design. Our goal is to develop methods to determine what underlies these interesting features and facilitate comparisons between how the cochlea works in different species or at different locations.

For single-partition box representations of the cochlea, information regarding how it works is entirely encoded in two mechanistic variables: differential pressure wavenumber, $k(x,\w)$, which encodes propagation properties, and the Organ of Corti effective impedance, $Z(x,\w)$, which encodes dynamics and may be used to determine effective dynamic representations of the Organ of Corti. The wavenumber and impedance are a window into properties such as effective stiffness, positive and negative damping, amplifier profile, incremental wavelengths, gain and decay, phase and group velocities, travel times, and dispersivity. Estimating these two mechanistic variables, $Z, k$, is therefore key to understanding what underlies cochlear features of high gain and place-based wavelet analyzers.

The wavenumber and impedance cannot be directly observed, and instead must be estimated from measurements. Here, we estimate the mechanistic variables (a) from observed variables (b) via utilizing model assumptions. As for (a), we use noninvasive response characteristics, such as the quality factors of psychophysical tuning curves. For (b), we leverage a previously developed model of the healthy mammalian cochlea \cite{paperA}. The model builds on classical box representations of the cochlea and is based on physics and phenomenon. It is valid near the peak region for low stimulus levels due to assumptions of linearity and short wave approximations. 

\begin{figure}[htbp]
    \centering
    \includegraphics[scale = 0.7]{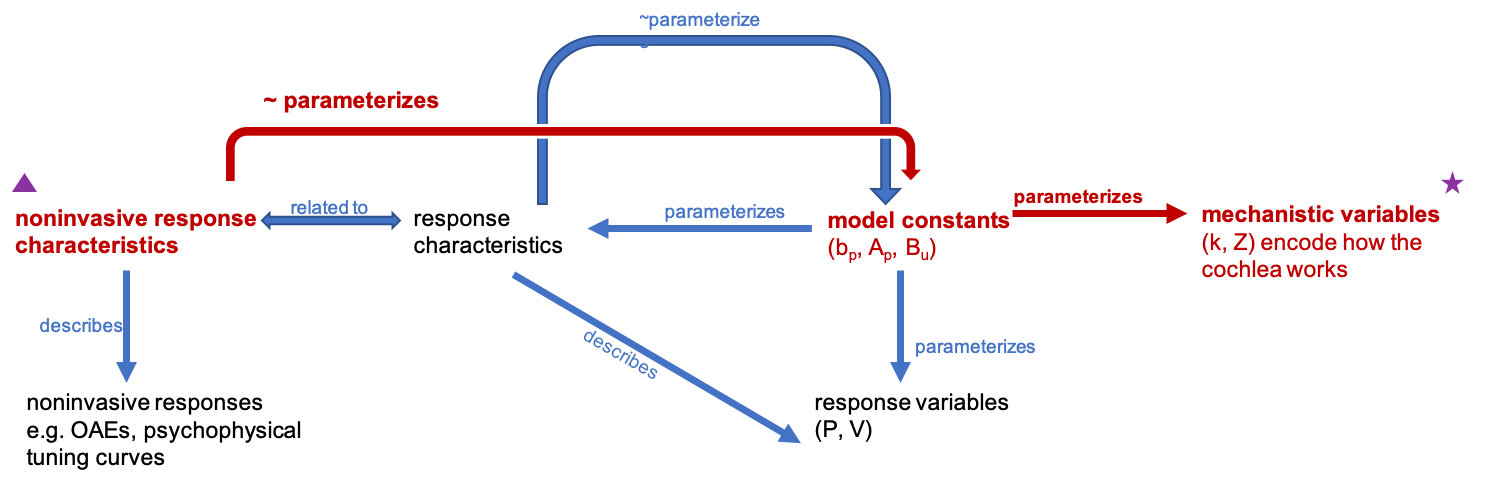}
    \caption{We develop methods for estimating mechanistic variables which encode how a box representation of the cochlea works (purple star) from noninvasive response characteristics (purple triangle). To do so, we leverage an analytic model of the healthy mammalian cochlea in which a set of three model constants ($\bp, \Ap, \Bu$) parameterizes the model mechanistic variables, $k(x,\w), Z(x,\w)$, as well as the model macromechanical response variables, $P(x,\w), V(x,\w)$. The magnitude and phase of the response variables may be described by filter characteristics such as peak frequencies, quality factors, and group delays that are reported from invasive experiments in some animals. Our objective (following the red text and arrows from the purple triangle to the purple star) is to develop methods for estimating the mechanistic variables from characteristics of noninvasive responses such as quality factors of psychophysical tuning curves. To do so, we first derive a parameterization of model constants in terms of noninvasive response characteristics by deriving then `inverting' expressions for noninvasive response characteristics in terms of model constants. For species where invasive measurements are reported, we may alternatively parameterize the model constants in terms of (invasive) response characteristics such as the quality factors and group delays of experimental mechanical (or neural) measurements that provide both magnitude and phase information.}
    \label{fig:schematic}
\end{figure}

In this paper, we first briefly summarize the model expressions relevant for our mechanistic study. We then use our model towards the goal of understanding how the cochlea workds by achieving the following objectives:

\begin{enumerate}
    \item  Develop methods to determine estimates for the mechanistic variables - wavenumber and impedance, from noninvasive response characteristics - specifically quality factors of psychophysical tuning curves. The paradigm is schematized in figure \ref{fig:schematic}.
    \item Provide estimates for mechanistic variables for humans. No previous estimates exist for humans due to the invasive surgical nature of experiments to obtain mechanical measurements from the cochlea.
    \item Compare the mechanistic variables between the human base and the human apex.
\end{enumerate}

\section{Summary of Model Expressions}

In order to develop the mechanistic variable estimation methods for understanding how the cochlear works, we leverage a previously developed interpretable analytic model of the mammalian cochlea \cite{paperA}. The analytic model was tested against existing data from various regions and species generated by experimental labs - e.g. \cite{temchin2005wiener}. The model has interpretable closed-form expressions for all variables (pressure, velocity, wavenumber, and impedance), and is parameterized by a small number of model constants. The model is appropriate for developing the estimation methods towards understandings how the cochlea works. This is the case because the aforementioned model properties enable us to derive expressions to compute model constant values from values of response characteristics such as bandwidths and maximum group delays of pressure or velocity \cite{paperA, paperB, thesis} as will be elaborated upon later in this paper. Building on these expressions, we may further develop methods to estimate values of model constants (and from them, mechanistic variables) from characteristics of noninvasive responses (such as psychophysical tuning curves and otoacoustic emissions). In this section, we provide model expressions for the mechanistic variables, $k,Z$.

The frequency domain representation of the model is in terms of $x, \w$. For the wavenumber, we collapse the dependence onto a single independent variable which is a normalized frequency or transformed space, $\B$,
\begin{equation}
\B(x, \w) \triangleq \frac{f}{\text{CF}(x)} \;,
\end{equation}

where CF$(x) = \text{CF}_{max}\exp{-x/l}$ is the known characteristic frequency map of the species. 

Here we summarize our model expressions for the mechanistic variables, wavenumber $\kx(\B)$, and effective impedance $Z(x, \w)$, which encode this family of cochlear representations work. Details of the model derivation based on physics and phenomenon, model testing, and model expressions for mechanistic variables (wavenumber and effective impedance) and response variables (pressure and velocity) were previously developed \cite{paperA}. In developing the model, we have not assumed any resonances or any other particular forms of impedance. We utilized qualitative information from Weiner-Kernel based estimates of the wavenumber from chinchillas \cite{shera2007laser}. We have further imposed purely forward traveling waves, and that the differential pressure traveling wave does not grow beyond its peak. The model expression we constructed for the wavenumber, $\kx$, is 
\begin{equation}
\kx(\B) \frac{l}{\B} = 2\Bu \frac{i\B+ \Ap}{(i\B-p)(i\B-\pconj)}\;.
\label{eq:kxfull}
\end{equation}

The expression for $\kx$ is closed form and easily interpretable. Its dependence on $x,\w$ is encoded in its dependence on $\B$ alone,  which couples the dependence on $x$ with the dependence on $\w$. Function-wise, this intriguingly couples the inhomogeneity (spatial variation of material properties) with dispersivity (separation of different frequency components of traveling waves as they propagate) in the cochlea. The assumption of scaling symmetry of $\kx$ is valid locally as the parameter values vary relatively slowly along the length of the cochlea \cite{paperA, thesis}.

The expression for $k$ is a rational transfer function that has a pair of complex conjugate poles, $p = i\bp - \Ap$ and  $\pconj = -i\bp - \Ap$, as well as a real zero at $i\B = -\Ap$. The three model constants, $\Ap, \bp, \Bu$ take on positive real values. The constant, $l$ is the space constant of the cochlear map, $\text{CF}(x)$, that is empirically known for a variety of species, including humans. 

The closed-form model expression for effective impedance, $Z$, is derived from the above expression for $\kx$, as well as the relationship between $\kx$ and $Z$ for the short-wave approximation of box model representations of the cochlea,

\begin{equation}
    Z(x,\w) \kx(\B(x,\w)) = -2i \r \w\;,
    \label{eq:ZfromK}
\end{equation}

where $\r$ is the density of scala fluid. Much can be inferred regarding how the cochlea works, \textit{qualitatively}, based on the form of the above expressions for $k$ and $Z$ and their variation as a function of $\B$ or $x,\w$ \cite{thesis}. However, in this paper, we focus on developing methods for \textit{quantitative} analysis as is relevant for comparative studies across species and locations. 

We previously also provided expressions for our model macromechanical responses, pressure and velocity, as parameterized by our three model constants \cite{paperA}. The responses (pressure or velocity), relative to the input at the stapes, can be characterized by a variety of frequency domain measures that describe their behavior - such as peak frequency, bandwidth, quality factor, group delay, and phase accumulation. We refer to these collectively as response characteristics. 

\section{Methods for Estimating Model from Macromechanical Response Characteristics}

Given values of the model constants, we can determine the mechanistic variables using equations \ref{eq:kxfull} and \ref{eq:ZfromK}. Our goal is to estimate mechanistic variables from noninvasive response characteristics which involves, as schematized in figure \ref{fig:schematic}, estimating model constant values from noninvasive response characteristics. In this section, we derive expressions for the three model constants in terms of (invasive) response characteristics, which is necessary towards ultimately estimating model constants (and from them, mechanistic variables) from noninvasive response characteristics.

It is desirable to parameterize the model constants in terms of response characteristics rather than determining them using responses themselves which suffer from incompleteness and noise issues. Furthermore, response characteristics are more readily available from the literature. Note that, to first approximation, the peak-centric response characteristics of pressure and velocity are similar for the parameter values of interest, and hence we do not distinguish between them here. 

In order to derive approximate expressions for model constants in terms of reported response characteristics, we first derive expressions for response characteristics parameterized by the three model constants  - schematized in figure \ref{fig:schematic}, then `invert' them. We deal with the following set of response characteristics: peak normalized frequency, $\B_{peak}$, group delay at the peak in the normalized frequency domain in cycles, $N$, and n dB bandwidth in the normalized frequency domain, $\text{BW}_{\text{n dB}}$. Note that $\text{BW}_{\text{n dB}}$ can easily be converted to the $f$ domain bandwidth by multiplication by CF$(x)$, and $N$ can be converted to the $f$ domain by division by CF$(x)$. 

We arrive at the following simple closed-form expressions for response characteristics in terms of model constants. The are especially valid for sharp filters which occur in the human base and human apex as well as the chinchilla base.

\begin{align}
    \B_{peak} = \bp \\
    N = \frac{\Bu}{2\pi\Ap}\\
    \text{BW}_{\text{n dB}} = 2\Ap \sqrt{\exp{\frac{\log(10) n}{10 \Bu}}-1} 
\end{align}

We then `invert' these expressions to arrive at the following formulas for the model constants in terms of the response characteristics.

\begin{enumerate}
    \item As $f_{peak} = \text{CF}(x)$, we generally set $\bp = 1$
    \item Solve for $\Bu$ from $N \text{ x } \text{BW}_{\text{n dB}} = \frac{\Bu}{\pi}\sqrt{\exp{\frac{\log(10) n}{10 \Bu}}-1} $
    \item Compute $\Ap$ from $\Bu$ determined above and either (a) the equation for $N$, as $\Ap = \frac{\Bu}{2\pi N}$, or (b) the equation for $\text{BW}_{\text{n dB}}$, as $\Ap = \frac{ \text{BW}_{\text{n dB}}}{2 \sqrt{\exp{\frac{\log(10) n}{10 \Bu}}-1}}$.
    \item Plug the estimated values for model constants computed above into the model expressions for wavenumber and impedance (equations \ref{eq:kxfull} and \ref{eq:ZfromK}) to study how the cochlea works
\end{enumerate}

\section{Methods for Estimating Model from Noninvasive Response Characteristics}

In this section, we develop methods to determine model constants, $\Ap, \Bu$ (and from them mechanistic variables $k,Z$ which encode how the cochlea works) from \textit{noninvasive} response characteristics. The method for determining the model mechanistic variables described in the previous section require response characteristics from invasive measurements, and can readily be applied to study mechanisms in animals for which we have invasive response characteristics. However, this is not feasible in humans with today's technology. Consequently, we must develop a way to determine the model constants and mechanistic variables from characteristics of noninvasive responses. Specifically, we may use quality factors of psychophysical tuning curves and/or group delays of stimulus frequency otoacoustic emission (SFOAE).

In order to achieve our goal and use these noninvasive response characteristics to determine the values for model constant needed to determine $k, Z$, we must first determine the relationship between the noninvasive response characteristics and the (invasive) response characteristics described above. This allows us to then use the methods outlined in the previous section towards our goal. Here we describe one such mapping which utilizes quality factors of psychophysical tuning curves and an assumed species invariant ratio.

Reported values of quality factors, $Q_{erb}$, are based on equivalent rectangular bandwidths. In contrast, our simple closed-form expression for bandwidth, and hence associated quality factor, $Q_n$, is for n dB bandwidths. To convert between $Q_{erb}$ and $Q_n$, we use an empirical relationship between these two measures for $n=10$. This relationship, $Q_{erb} = \alpha Q_{10}$, where $\alpha$ is in the range $1.7-1.8$, was obtained from ANF tuning curves and found to be largely species and CF invariant \cite{shera2003stimulus}. Note that $Q_{10}$ is the same whether defined in $\B$ or $f$ domains, and that $\frac{1}{Q_{10}} = BW_{\text{10 dB}}$ (bandwidth defined in the $\B$ domain) because $\B_{peak} = 1$. This allows us to replace the second step in the previous section (determining $\Bu$) with,

\begin{equation}
\frac{N}{Q_{erb}} = \frac{\Bu}{\alpha \pi}\sqrt{10^{\frac{1}{\Bu}}-1} \;.
\end{equation}

Consequently, using the equation above,  we require values for $\frac{N}{Q_{erb}}$ in order to estimate $\Bu$. To do this, we use the ratio $g = \frac{Q_{erb}}{N}$ which was empirically found to be a constant in chinchilla ($g \approx 1.25$) and assume that this ratio is species-independent for most of the length of the cochlea - note that the tuning ratio $r = \frac{Q_{erb}}{N_{sfoae}}$ has been shown to be species invariant \cite{triangle}. Using these values for $g$ and $\alpha$, we solve for $\Bu$. This gives $\Bu \approx 7$ for various species across the length of the cochlea where the sharp-filter approximation holds.

Now that we have determined $\Bu$, we may now determine $\Ap(\text{CF(x)})$ from reported values of quality factors of psychophysical tuning curves. Specifically, following step 3-(b) of the previous section, we use reported $Q_{psych-erb}(\text{CF(x)})$ and the $\Bu$ determined above. We assume that the quality factors of psychophysical tuning curves approximate the quality factors of macromechanical responses, $Q_{psych-erb} \approx Q_{erb}$, and use $\alpha$ to substitute  $\text{BW}_{\text{10 dB}}$ of step 3-(b) with $\text{BW}_{\text{10 dB}} = \frac{\alpha}{Q_{erb}}$. 

\begin{equation}
    \Ap(\text{CF}) = \frac{\alpha}{2 Q_{psych-erb}(\text{CF}) \sqrt{ 10^{\frac{1}{\Bu}} - 1}}
\end{equation}

Another way to estimate $\Ap(\text{CF})$ from quality factors of psychophysical tuning curves, is by using step 3-(a), and substituting $N = \frac{Q_{erb}}{g}$,
\begin{equation}
    \Ap(\text{CF}) = \frac{g \Bu}{2 \pi Q_{psych-erb}(\text{CF})}
\end{equation}

Using either expression above, we use reported quality factors of psychophysical tuning curves from humans to determine $\Ap$. Both equations yield very similar estimates for $\Ap(\text{CF})$ from $Q_{psych-erb}(\text{CF})$. Note that we may alternatively have estimated $\Ap(\text{CF(x)})$ using another noninvasive response characteristic which can be measured in humans. Specifically, we may determine $\Ap$ using step 3-(a) with group delays of stimulus frequency otoacoustic emissions if we make assumptions regarding the relationship between $N$ and $N_{sfoae}$. This relationship has been the subject of previous studies \cite{shera2008delay2delay}. 

With this we have achieved our first objective of developing methods for estimating model constant values (and thereby mechanistic variables) from noninvasive response characteristics. We apply these methods to two points in the human cochlea which have the following reported $Q_{psych-erb}$ determined using a non-simultaneous masking paradigm with low stimulus levels \cite{shera2002revised}: at 10 kHz, $Q_{psych-erb} \approx 25.34$ and at 1 kHz, $Q_{psych-erb} \approx 12.7$. Using the equations above, these result in $\Ap \approx 0.055$ and $\Ap \approx 0.11$ respectively.

\section{Estimates for Human Wavenumber and Impedance}

\begin{figure}[htbp!]
     \centering
     \begin{subfigure}
         \centering
       \includegraphics[scale = 0.7, trim={0 10.4cm 0 0}, clip]{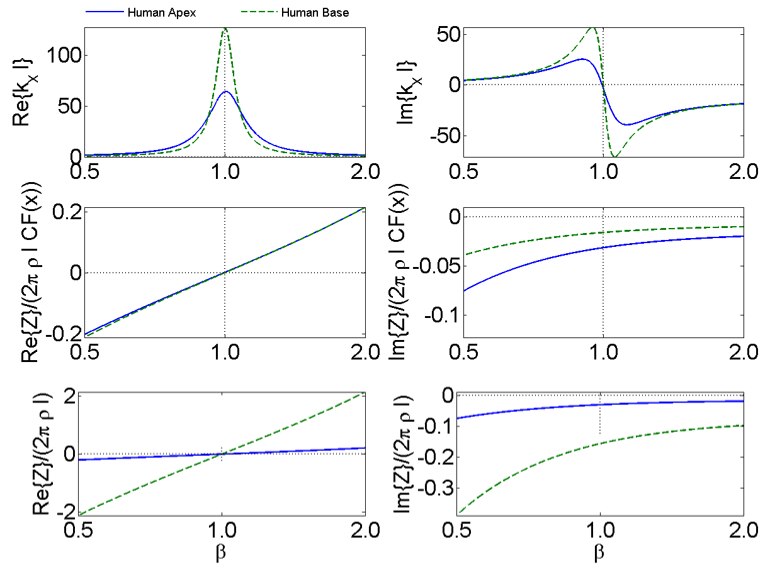}
     \end{subfigure}
     \begin{subfigure}
         \centering
        \includegraphics[scale = 0.7, trim={0 0 0 10cm}, clip]{figures/human}
     \end{subfigure}
        \caption[Estimated wavenumbers and impedances in the human base and apex]{Estimated wavenumbers and impedances in the human base and apex: The top panel shows the real and imaginary parts of the model generated human wavenumber, and the bottom panel shows the real and imaginary parts of the impedance. The blue solid lines are for a point in the apex (CF = 1 kHz, where $Q_{psych-erb} \approx 12.7$) which has estimated model constants $\Ap = 0.11, \Bu = 7$, and the dashed green lines are for a point in the base (CF = 10 kHz, where $Q_{psych-erb} \approx 25.34$) which has estimated model constants $\Ap = 0.055, \Bu = 7$. The values for $\Ap, \Bu$ are estimated from noninvasive response characteristics in the previous section. }
        \label{fig:human}
\end{figure}

With the method introduced above for determining model constant values from noninvasive response characteristics, we can now (2) determine the wavenumber and impedance from human noninvasive response characteristics, and also (3) differentiate between their profiles in the human base and apex (which provides us with information regarding how these two regions of the cochlea function differently). We do so by first plotting (in figure \ref{fig:human}) the wavenumber and impedance of two points along the length of the cochlea - one from the apex with a low CF, 1 kHz, and another from the base with a higher CF, 10 kHz. We arrive at the mechanistic variables in the plot using the two sets of model constants estimated in the previous section from noninvasive characteristics (basal location $\Ap = 0.055, \Bu = 7$, and apical location $\Ap = 0.11, \Bu = 7)$.

As seen in figure \ref{fig:human}, the real part of the wavenumber for a particular frequency, $f$, has a peak, the value of which determines the minimum incremental wavelength of the differential pressure traveling wave as it propagates along the length of the cochlea. The peak is greater in the base than the apex, indicating that the wavelength is much smaller in the base than in the apex near the location at which the wave peaks (though this is not the case outside the peak region). The imaginary part of the wavenumber has a peak and trough that are larger in the base than the apex which indicates that the gain and dissipation accumulate more quickly about the peak of the differential pressure traveling wave in the base than in the apex. This can alternatively be observed in the corresponding pressure response.

The imaginary part of the impedance at a particular location, $x$, is negative (indicating an effective stiffness rather than mass). The effective stiffness is greater in the base than in the apex. If we interpret the effective impedance, $Z$, as a \textit{local} impedance due to properties at a single point, we may infer that the stiffness of the Organ of Corti is greater in the base than in the apex. The real part of the impedance is negative prior to the peak frequency at that point (indicating an active component and power flux into the traveling wave for single partition box representations) then positive after it (indicating absorption). The effective positive and negative damping are greater in the base than the apex. However, $\Re\{Z\}$ normalized by CF$(x)$ is is approximately equal in both the base and apex, suggesting that certain properties of the mechanisms are retained across location. Building on $\Re\{Z\}$ and other model variables, it is of interest to study the mechanistic variables in the context of power flux into the traveling wave in the single partition box model framework.

While no previous estimates for human mechanistic variables are available, existing literature does provide us with estimates of the impedance (or proxies for it) at single locations or frequencies for certain species \cite{dong2013detection, de1999inverse}. In addition to anticipated quantitative differences between the various estimates for $Z$ due to differences in species and locations, there are some qualitative differences: For instance, \cite{de1999inverse} has a $\Re\{Z\}$ that, about the peak, is negative then positive and crosses zero near the peak (which is qualitatively consistent with our model) but it differs from our estimates in terms of where the zero crossing occurs, and the behavior of $\Re\{Z\}$ away from it. Our $\Im\{Z\}$ is negative for all frequencies and the previous studies are consistent with this for the most critical frequencies or locations.

\section{Contributions and Future Work}

In this paper, we developed methods to estimate mechanistic variable which will enable us to study how the cochlea operates in various species, regions, or pathologies that have different response characteristics but the same underlying model structure, and determine differences between these various cases. The model utilized in developing the estimation method is for healthy mammalian cochleas at low levels. The estimation methods are best suited for cases where the tuning is relatively sharp as is the case in the human base and apex and chinchilla base.

We used the estimation methods to determine human wavenumber and impedance - for which there have been no previous estimates, from reported values of noninvasive response characteristics. We also discussed differences in these mechanistic variables between the base and the apex. From these estimates, it is possible to derive other mechanistic information, such as phase velocity, dispersivity, power flux, effective stiffness, as well as their differences between the base and the apex to develop a deeper understanding of function. 

We may also extend the methods to derive mechanistic information from another noninvasive response characteristic - specifically stimulus frequency emission group delays. In addition to comparative analysis built on estimating values of model constants from characteristics, we can study the model \textit{expressions} for wavenumber and impedance more deeply (without specifying model constant values) as they provides us with a general understanding of how the cochlea functions across mammalian species. In addition to their use for studying auditory physics, the model and estimation methods are also useful for developing easily tunable auditory filters for engineering applications and perceptual studies \cite{paperB}.

\bibliography{references}

\end{document}